\documentstyle[12pt]{article}
        \def\be{\begin{equation}}
        \def\ee{\end{equation}}

\begin{document}
\begin{center}
{\bf A Multi-Species Asymmetric
Simple Exclusion Process\\ and its Relation to Traffic Flow}

\vspace {1cm}

{V. Karimipour} \\
\vspace {1cm}
School of Physics,\\
Sharif University of Technology\\
P.O.Box 11365-9161\\
and\\
Institute for Studies in Theoretical Physics and Mathematics\\
P.O.Box 19395-5746\\
Tehran, Iran\\
\vspace {1cm}

{Abstract} \\
\end{center}
Using the matrix product formalism we formulate a natural p-species generalization of the asymmetric simple
exclusion process. In this model particles
hop with their own specific rate and fast particles can overtake slow ones
with a rate equal to their relative speed.
We obtain the algebraic structure and study the properties of the
representations in detail. The uncorrelated steady state for the open system is obtained and in the ( $ p\longrightarrow \infty)$ limit, the
dependence of its characteristics on the distribution of velocities is
determined. It is shown that when the total arrival rate of particles exceeds
a certain value, the density of the slowest particles rises abroptly.

\vspace {1 cm}
\newpage
\section{Introduction}

The Matrix Product Ansatz (MPA), first introduced in the work of Hakim
and Nadal [1],become popular in the study of one dimensional stochastic
exclusion models
after Derrida et al [2] applied this technique to the simplest such
model,namely the totally asymmetric simple exclusion process or ASEP[3].
For a recent review see [4]. They were able to find nontrivial representations for the relevant algebra
and calculate many physical properties of this process. Since then this
technique has been applied to many other interesting stochastic systems,
both with periodic and with open boundary conditions. Janowsky and Lebowitz
[5] have considered the case where there is a single impurity (a fixed
blockage) on the lattice, where hopping rates are reduced, and have found that
this single blockage has global effects, due to long range correlations.
The case of a moving blockage and the formation of
shocks has also been addressed in several works [6-7].
A model consisting of two species moving in opposite directions have been studied in Evans et al [8]
where spontenous symmetry breaking has been observed.
Finally MPA has been formulated for models with quasi parallel updatings
[9-11].\\
In most of these works , a simplified model of traffic flow [12] is cited
as a physical application, specially when impurities are present which
when encountered decrease the rate of hopping of other particles [13].
Evans [14] has applied MPA to the case of several kinds of particles
hopping with their own rates on a ring and has found that above a critical
density
a macroscopic number of holes are condensed in front of the slowest particle.
In the work of Evans, however, particles can not overtake, and the
order of particles is unchanged during the process.\\

Turning to the mathematical aspect of the problem, a matrix product state
can be understood as a generalization of ordinary factorizable states
with a product measure in which numbers are replaced with non-commuting
objects.This allows the original problem to be reformulated in terms of
algebraic relations. The main advantage of this thecnique is that once
a nontrivial representation is found, all the physical quantities like
the currents , densities and correlation functions can in principle be obtained
from the representation. This is of course not a simple task and one should
overcome many difficulties of combinatorial type in calculating explicit
form of the matrix elements of product of operators.\\
Whether or not the zero energy state of a Hamiltonian with nearest neighbor
interaction can always be formulated as a matrix product state has been
answered in the affirmative by Sandow [15].
Thus to any Hamiltonian of the above type, corresponds an algebra in the MPA
formalism. However finding nontrivial representions of this algebra may be
harder than the original problem itself. Therefore some authors [16,17] have
tried to reverse the problem, thats to begin with a quadratic algebra,with
hopefully simple representations and then to find the relevant stochastic
process. In [17] the algebras considered are appearantly  related to the
partially assymetric case. This restriction has a two-fold disadvantage .
First they obtain the concrete form of the algebraic relations only in
cases for low number of species, and second they
do not obtain [17] " a general framework for arbitrary N ( number of species ),
which when specialized to N=2 gives the know results."\\
Owr aim in this paper is to start from a suitable algebra and proceed to find
a natural generalization of the one species ASEP.We will show that such
a natural generalization exists. We will obtain the algebra,
its representations and
many of its properties. In this p-species ASEP each particle of type j, hops
with rate $ v_j$ to its right neighbor site , and when it encounters a particle
of type $ i $ with $ v_j > v_i $ they interchanges their site with rate $
v_j - v_i $, as if the fast particle overtakes the slow one. In any other case
the attempted move of the original particle is rejected. The model seems to be
relevant as a simplified model of one way traffic flow. We consider open
boundary
conditions where particles ( or cars ) enter the system ( the highway ) with a
rate proportional to their speed and leave at the other side.\\
The paper is organized as follows: In section 2 we introduce the algebra and
solve its'consistency conditions.
In section 3 we obtain the Hamiltonian
which corresponds to this algebra .
In section 4 we obtain the uncorrelated
steady state and finally in section 5 we construct the infinite dimensioal representations of the algebra and present some of its useful properties.
In all the steps we compare our resulsts with the one
species ASEP.   \\

\section{The Algebra}

We begin with an algebra generated by the elements $D_i, i=1,...p $
and $E$, with the following relations:
\begin{eqnarray} D_i E &=& {1\over v_i} D_i + E \\
 D_j D_i &=& \xi_{ji} D_j + \eta_{ji} D_i \hskip 1cm j>i\end{eqnarray}
where the parameteres $ v_i$ are finite real numbers and the parameters
$\xi_{ji}$ and$\eta_{ji}$ are to
be determined. In writing (1,2) we have had in mind a totally assymetric
exclusion process.In order to be consistent with associativity we have
to check that
\be (D_j D_i)E = D_j(D_iE)\ee
Using (1) and (2), this requirement determines $\xi_{ji}$ and $\eta_{ji}$ as
\be \xi_{ji} = { v_i \over{v_i - v_j }} \hskip 2cm \eta_{ji} =
{ - v_j \over{v_i - v_j }}\ee
For the present we take all the $v_i$ to be different. Later we will relax this
condition. We should also check that
\be (D_k D_j)D_i = D_k(D_j D_i ) \hskip 1cm k >j >i \ee
Using (2) we find that this relation imposes no new relations on the parameters.
Since any monomials of the form $D_k D_j...D_i $ with $ k >j >...i$ can
be reduced to
a linear combination of generators, associativity is guaranteed for all the
monomials of higher degree.
The final form of the algebra reads
\begin{eqnarray} D_i E &=& {1\over v_i} D_i + E \\
 D_j D_i &=& { 1 \over{v_i - v_j }}( v_i D_j - v_j D_i ) \ \ \ \ \ \  j>i \end{eqnarray}
where the parameters $ v_1 .... v_p $ are free.  \\

\section{ The Hamiltonian and the Process}

We consider a one dimensional chain of N sites. The Hilbert space of each site
is $ p+1 $ dimensional. The Hamiltonian is written as:
\be H = h^1 + H^B + H^N = h^1 + \sum_{k= 1}^{k=N-1}h^B_{k,k+1} + h^N \ee
Where $ H^B$ is the bulk Hamiltonian and $ h^1 $ and $h^N$ are boundary terms.
In the matrix product formalism, the steady state of this Hamiltonian can be
written as:
\be \vert P > = < W \vert {\cal D}\otimes{\cal D}\otimes {\cal D}\otimes
....{\cal D}\otimes{\cal D}\vert V>\ee
where
$ {\cal D } = \left( \begin{array}{l}E\\D_1\\.\\.\\D_p\end{array}\right)$
and $ \vert V>$ and $< W\vert $ are vectors in an auxiliary space.
According to MPA, we have $ H\vert P > = 0 $ if the following relations are
satisfied:
\be h^B {\cal D}\otimes {\cal D} = X \otimes {\cal D} - {\cal D}\otimes X \ee
\be (h^N{\cal D} - X) \vert V> = 0 \ee
\be <W\vert (h^1{\cal D} + X) =0\ee
where we take $ X $ to be $ X = \left( \begin{array}{l}{-1}\\x_1\\.\\.\\
x_p\end{array}\right)$ with $ x_i $ as c-numbers.
Anticipating the process from the form of the algebra,we write the bulk
Hamiltonian as follows:
\be h^B = -\sum _{i=1}^p y_i ( E_{0i}\otimes E_{i0} - E_{ii}\otimes E_{00})
- \sum _{j>i}^p y_{ij} (E_{ij}\otimes E_{ji} - E_{jj}\otimes E_{ii})\ee
Here the matrices $ E_{ij}$ act on the Hilbert space of one site and have
the standard definition: $ (E_{ij})_{k,l} = \delta_{ik}\delta_{jl}$.
The first term expresses an ASEP for each species of particles (i) with
rate $ y_i $. The second term represents an interchange of particles of type
$ (j > i) $, with rate $ y_{ij} $. The parameters $y_i$ and $ y_{ij} $ are
to be determined. \\

The natural choice of the boundary terms should be such that particles of
type $ (i) $ are injected at the left and extracted at the right with
their specific rates say $ \alpha_i $ and $ \gamma_i $ respectively.
So we take:
\be h^1 = -\sum _{i=1}^p \alpha_i ( E_{i0} - E_{00})\ee
\be h^N = -\sum _{i=1}^p \gamma_i ( E_{0i} - E_{ii})\ee
Inserting (13) in (10) leads to the following equations:
\begin{eqnarray} {y_i\over p} D_i E &=& {D_i \over p } + x_i E \\
 {y_{ij}\over p} D_jD_i &=& x_j D_i - x_i D_j \end{eqnarray}
Comparision with (6) and (7) shows that:
\be x_i = {v_i\over p} \ \ \ y_i = v_i \ \ \ y_{ij} = v_j - v_i \ee
Inserting the values of the parameters $ y_i$ and $ y_{ij} $ in (13)
clearly highlights the physical process governed by the Hamiltonian.
Denoting a vacant site
by the symbol $ \phi $ and a site occupied by a particle of type i,
by the symbol
$ A_i $, the process defined by $H^B$ is :
\begin{eqnarray} A_i \phi &\longrightarrow &\phi A_i \ \ \ {\rm with\ \  rate }\ \ \ v_i\\
 A_j A_i &\longrightarrow & A_i A_j \ \ \ j>i  \ {\rm with\ \  rate}
\ \ \ v_j-v_i \end{eqnarray}
In order for all the rates to be positive we restrict the range of
$ v_i $'s as:
\be v_1\leq v_2 \leq v_3....\leq v_p \ee
Inserting (14) and (15) respectively in (11) and (12) we find the the following
explicit matrix equations:
\be \Bigg( \left( \begin{array}{lllllll} 0&{-\gamma_1}&{-\gamma_2}&{-\gamma_3}&.&.&{-\gamma_p}\\
0&{\gamma_1}&0&0&.&.&0\\0&0&{\gamma_2}&.&.&.&0\\0&0&0&{\gamma_3}&.&.&0\\.&.&.&.&.&.&.\\.&.&.&.&.&.&.\\
0&0&0&0&.&.&{\gamma_p}
\\ \end{array} \right)
\left( \begin{array}{l}E\\{D_1\over p}\\{D_2\over p}\\.\\.\\.\\{D_p\over p}\end{array}\right)
-
\left( \begin{array}{l}{-1}\\{v_1\over p}\\{v_2\over p}\\.\\.\\.\\{v_p\over p}\end{array}\right)
\Bigg)\vert V> = 0 \ee

\be < W \vert \Bigg( \left( \begin{array}{lllllll} {\alpha_1 +\alpha_2+ ...\alpha_p}&0&0&
.&.&.&0\\
{-\alpha_1}&0&0&.&.&.&0\\{-\alpha_2}&0&0&.&
.&.&0\\{-\alpha_3}&0&0&.&.&.&0\\.&.&.&.&.&.&.\\.&.&.&.&.&.&.\\{-\alpha_p}&0&0&.&.&.&
0\\ \end{array} \right)
\left( \begin{array}{l}E\\{D_1\over p}\\{D_2\over p}\\.\\.\\.\\{D_p\over p}\end{array}\right)
+
\left( \begin{array}{l}{-1}\\{v_1\over p}\\{v_2\over p}\\.\\.\\.\\{v_p\over p}\end{array}\right)
\Bigg) = 0 \ee
These equations then yield:
\be D_i \vert V> = {v_i\over \gamma_i} \vert V> \ee
\be < W \vert E =  < W\vert {v_i\over {p \alpha_i}} \ee
and
\be {{v_1 + v_2 + ... v_p}\over p}=1 \ee
Equation (26) which means that the average speed of the particles is
unity, is not really a condition on the rates, since by a rescaling of
time one can always obtain this condition.
Hereafter we will write this condition simply as $ < v > = 1 $.
Equation (25) however implies that $ {v_i \over {p \alpha_i}} $ is independent of i.
Thus we set $ \alpha_i ={1\over p} \alpha v_i $.\\
In order that equation (24) be consistent with (7), we require that :
\be \bigg( D_j D_i - { v_i D_j - v_j D_i\over{v_i - v_j }} \bigg) \vert V > = 0 \ee
After using (24) and rearranging terms, this condition shows that  $$ \gamma_i - v_i = \gamma_j - v_j .$$ Thus we set
$$\gamma_i = v_i + \beta - 1 $$                         \\
Therefore we have
\be D_i \vert V> = {v_i\over {v_i + \beta - 1 }} \vert V> \hskip 2cm
< W \vert E =  < W\vert {1\over \alpha} \ee
The meaning of the parameters $ \alpha $ and $ \beta $ are now clear. From (26)
we find:
$$ \alpha_1 + \alpha_2 + ... \alpha_p = \alpha \hskip 2cm { \gamma_1 + \gamma_2 + ... \gamma_p \over p } = \beta$$
Thus $ \alpha$ and $ \beta $ are respectively the total rate of injection
and the average rate of extraction of particles.
In order for the individual rates to be positive we restrict further
the range of
speeds as
\be 1-\beta \leq v_1\leq v_2 \leq v_3....\leq v_p \ee
The above results clearly justifies this model as a natural
generalization of the
1-species ASEP. \\
We conclude this section with formulas for the current operators. A simple
way to obtain these, is to directly refer to the process (19-20) and write
the equation for the one point function $ < \tau^{(i)}_k>$ which is the
average density of particles of type $(i)$ at site k. Determining from
(19-20) various ways of increasing and decreasing of this density and
denoting the probabilty of site $ k $ to be vacant by $ \epsilon_k $
we find:
\begin{eqnarray} {d\over dt}<\tau^{(i)}_k> &=& v_i<\tau^{(i)}_{k-1} \epsilon_k >
- v_i<\tau^{(i)}_{k} \epsilon_{k+1} >  \cr
&+&\sum_{j<i} (v_i - v_j) <\tau^{(i)}_{k-1} \tau^{(j)}_k >
- \sum_{j<i} (v_i - v_j) <\tau^{(i)}_{k} \tau^{(j)}_{k+1} >  \cr
&-& \sum_{j>i} (v_j - v_i) <\tau^{(j)}_{k-1} \tau^{(i)}_k >
+\sum_{j>i} (v_j - v_i) <\tau^{(j)}_{k} \tau^{(i)}_{k+1} > \end{eqnarray}
Note that inside any correclation function we have $ \epsilon_k = 1 - \sum_i \tau^{(i)}_k $.
Equation (30) can be rewritten as a continuity equation
$$ {d\over dt}< \tau^{(i)}_k > = < J^{(i)}_k >-<J^{(i)}_{k+1}>$$
where
\be <J^{(i)}_k> = v_i <\tau^{(i)}_{k-1} \epsilon_{k}> + \sum_{j<i} (v_i - v_j) <\tau^{(i)}_{k-1} \tau^{(j)}_{k} >
-\sum_{j>i} (v_j - v_i) <\tau^{(j)}_{k-1} \tau^{(i)}_{k} > \ee
According to MPA
$$ < J^{(i)}_k > = {<W\vert C^{k-2}J^{(i)} C^{N-k}\vert V>\over
<W\vert C^N\vert V>}$$
where $ C:= E+{1\over p } D := E+{1\over p } (D_1 + D_2 + ... D_p )$.
The current operators can be read from (31) to be:
\be J^{(i)} = v_i {D_i\over p} E + \sum_{j<i}(v_i - v_j){D_i\over p} {D_j\over p} - \sum_{j>i}(v_j - v_i ) {D_j\over p} {D_i\over p} \ee
Using eqs.(6,7), we obtain:
\begin{eqnarray}  J^{(i)} &=& {v_i\over p} ( {1\over v_i}D_i + E ) - {1\over p^2}\sum_{j<i}(v_j D_i - v_i D_j ) -
{1\over p^2} \sum _{j>i}(v_j D_i- v_iD_j )\cr
&=&{1\over p} D_i + {v_i \over p} E - {1\over p^2}( \sum_{j\ne i}v_j)D_i + {v_i\over p^2} (\sum_{j\ne i}D_j)\cr
&=& {1\over p}D_i + {1\over p}v_i E - {1\over p^2}( p - v_i ) D_i + {v_i\over p^2} ({D\over p^2} - D_i )\cr
&=& {v_i\over p} C \end{eqnarray}
We thus obtain
\be < J^{(i)}> = {v_i\over p}{<W\vert C^{N-1}\vert V>\over
<W\vert C^N\vert V>}\ee
Therefore all the currents are simply proportional to the average current $ J $, however
$ J $ itself has a highly nontrivial dependence on the hopping rates.  \\
In the next section we will find the one dimensional representations of
the algebra. This case corresponds to the steady state being characterized by
a Bernouli measure. Although very simple this steady state has a rather rich structure.

\section{One Dimensional Representations and the Uncorrelated Steady State }

In the one dimensional representation the operators $ D_i $ and $ E $ are
represented by c-numbers $ \Delta_i $ and $ e$ respectively. From equation (28)
we have :
\be \Delta_i = { v_i\over { v_i + \beta - 1 } } \ \ \ \ \ e = {1\over \alpha}\ee
combining (35) and (6) gives the condition $ \alpha + \beta = 1 $ on
the average rates which is of the same form as in the one species asep.
This, together with (29) means that for an uncorrelated steady state to exist,
the minimum speed $ v_1 $ should be greater than the average arrival rate
 $ \alpha$.
The steady state is now given by $ \vert P> = \vert \rho >^{\otimes N} $ where
\be \vert \rho> = {1\over c} \left( \begin{array}{l}e\\ {\Delta_1\over p}\\
{\Delta_2\over p} \\.\\ . \\ .\\ {\Delta_p\over p} \end{array}\right)\ \ \ \
c = e + {1\over p}(\Delta_1 + \Delta_2 ... \Delta_p) \equiv e +{1\over p} \Delta\ee
The density and current of particles of type $ (i) $ and the total density and
the total current are all site independent and are respectively given by (see 34)
\be \rho(\alpha,i) = {{\Delta_i\over p} \over {e + {\Delta\over p}}} \hskip 1cm  J(\alpha,i) = {{v_i\over p} \over
{e + {\Delta\over p}}} \hskip 1cm \rho(\alpha) = {{\Delta\over p} \over {e + {\Delta\over p}}}
\hskip 1cm J(\alpha) = {1 \over {e + {\Delta\over p}}} \ee
It is better to consider the limiting case $ p\longrightarrow \infty $,
that is,
we assume that particle speeds are taken from a continous probability
distribution $ P(v)$. Condition (26) is then transformed into
$$ < v > := \int v P(v)dv = 1 $$
Discrete quantities $ {1\over p} f(i) $ are transformed into the continous
functions $ f(v) P(v)$ and sums into integrals.
Instead of (36) we have
\be  \rho(\alpha,v)= {\Delta (\alpha,v)P(v) \over {e + \Delta(\alpha)}} \hskip 2cm  J(\alpha,v) = {v P(v) \over
{e + \Delta(\alpha)}}\ee \be \rho(\alpha)= {\Delta(\alpha) \over {e + \Delta(\alpha)}} \hskip 1cm J(\alpha) =
{1 \over {e + \Delta(\alpha)}} \ee
where
\be \Delta(\alpha,v) ={v\over { v-\alpha}} \ \ {\rm  and}\ \ \Delta(\alpha) = \int {v\over { v-\alpha}} P(v) dv \ee
As an example,in the following,we take a distribution
which at low speeds vanishes as some power of $ v-\alpha$, so that $ \alpha$
is indeed the minimum speed of particles. Requiring that the distribution
has an exponential decay rate $ \lambda $ at high speeds and the
average speed be
unity gives:
\be P_{\lambda}(v) ={1\over {\lambda ^{ m+1}\Gamma( m+1)}}(v-\alpha)^{m}
e^{-(v-\alpha)\over \lambda}\ee
where $ m+1 = {{1-\alpha}\over \lambda} $.Here $ m $ is not necessarily
an integer.
A similar choice of $ P(v)$ has also been made in [14].Note that since
$ m$ is to
be positive for each choice of $ \lambda$, we should have
$ 0< \alpha< 1 - \lambda$.
This distribution is peaked at $ v = 1-\lambda $.
Inserting (41) into (40) and doing the integral gives:
\be \Delta(\alpha)={{\lambda-1}\over{\lambda+\alpha-1}}\ee
from which we obtain
\be J(\alpha)={\alpha(\lambda+\alpha-1)\over {\lambda-1+ \alpha \lambda}}
\hskip 2cm  \rho(\alpha)={\alpha(\lambda-1) \over {\lambda-1+ \alpha
\lambda}}\ee
Eliminating $ \alpha $ between the above equations gives the current versus
density:
\be J(\rho)=\rho\Bigg(1- {\rho\over { 1- \lambda( 1-\rho)}}\Bigg)\ee
The curves $ J( \alpha)$ and $ J(\rho ) $ are shown in figs.(1 and 2) for various
values of the widths of the distributions. Note that for zero width ( $ \lambda=0 $)
the familiar results of the one species ASEP are obtained, namely:
$ \rho = \alpha $ and $ J = \rho(1 - \rho )$. For any finite value of $ \lambda $
$ \rho $ is an increasing function of $ \alpha $ but $ J( \alpha ) $ has a
maximum at ${ \tilde \alpha}:= { {1-\lambda}\over \lambda}(1-\sqrt{1- \lambda})$
The current at this optimal value of the arrival rate is
\be J_{max}(\lambda) = (1-\lambda)\Bigg({1-\sqrt{1- \lambda}\over \lambda}\Bigg)^2\ee
The maximum of $ J $ as a function of $\rho $ occurs at ${\tilde \rho}=
{1\over \lambda}\Big({\sqrt{1-\lambda}}-(1-\lambda)\Big)$.
It is seen that the maximum current has it is highest value of $ 1\over 4 $ only
when all the particles have the same hopping rates. It is interesting to
note that
even in such a simple uncorrleated steady state, a variance in hopping rates
reduces the current. \\

The final quantity we consider is the average density of particles of different
speeds,$ \rho(\alpha ,v ) $ as a function of $ v $. From (38-41) we have
\be \rho (\alpha,v) \propto v (v-\alpha)^{ m-1}
    e^{-(v-\alpha)\over \lambda}\ee
The interesting point is that when $ m\geq 1$ ( or $ \alpha \leq 1-2\lambda $ )
the density vanishes at the lowest speed $ \alpha$ and has a maximum at
$ v_{max} := {1\over 2} ( 1 - \lambda + \sqrt{(1-\lambda)^2-4\lambda \alpha})$.
However when $ m<1 $ or $ (\alpha > 1-2\lambda )$ the distribution changes
and the density of the slowest particles rises abroptly .
This transition is one of the interesting features of this process and is somehow reminicent of
the Bose condensation first noted in [14].

\section{Representations}
As in the one-species ASEP the representations of this algebra are either one dimensional or infinite dimensional.
To see this we first note that there is no non-zero eigenstate of $ E $,say
$ \vert e> $ such
taht $ E\vert e > = {1\over v_i } \vert e > $ for any $ i $. Since if there is
any such vector then by acting on both sides with $ D_i $ and using
(5) we obtain
$ {1\over v_i} \vert e> = 0 $ and since $ v_i $ is finite this means
that $ \vert
e > = 0 $. Following [2] we know show that in any finite dimensional
representation
of the algebra all the generators mutually commute. For if the
representation is
finite dimensional then the matrices $ E-{1\over v_1},... E-{1\over v_p}$
having no
zero eigenvalue, are all invertible and hence (5) gives $ D_i = E
(E-{1\over v_i})^{-1}$ which in turn means that the representation is
commutative.
Thus the representations of this algebra are either one dimensional or infinite
dimensional. Note that in the one dimensioanl case the expression
$D_i = E (E-{1\over v_i})^{-1}$ automatically satisfy (6,7). Thus such one
dimensional
representations really exist, and were used in section (4) to find the
uncorrelated steady state.\\
To find the infinite dimensional representations we assume that there is one
single vector denoted by $ \vert 0 > $ such that
\be D_i \vert 0> = d_i \vert 0 > \hskip 1cm i= 1,...p \ee
where the paramters $ d_i $ are to be determined.
We then consider the vector space $ W $ spanned by the formal vectors
$\{ \vert n > := E^n \vert 0 > , n= 0,1,2,....\}$.Clearly
\be E\vert n > = \vert n+1> , \hskip 1cm \forall n \ee
Iterating (5) we find
\be D_i E^n = v_i^{-n}D_i + v_i^{-n+1}E +v_i^{-n+2}E^2 + ... v_i^{-1}
E^{n-1} + E^n \ee
Thus we obtain:
\be D_i\vert n > = v_i^{-n}d_i\vert 0 > + v_i^{-n+1}\vert 1> + v_i^{-n+2}
\vert 2> + ... v_i^{-1}\vert n-1> + \vert n> \ee
In order to chech eq.(6) in this representation it is enough to consider only
the state $ \vert 0 > $, since all the other states are obtained by acting
on this
state using the algebraic relations which have been found to be consistent.
Therefore we require that
\be \bigg( D_j D_i - { 1 \over{v_i - v_j }}( v_i D_j - v_j D_i \bigg) \vert 0 > = 0 \ee
this fixes the parameters $ d_i $ to be $ d_i = { v_i \over {\epsilon + v_i }}$
where $ \epsilon $ is a constant independent of $ i $ . The explicit matrix
form of the generators are then as follows:
\be E = \left( \begin{array}{lllllll} 0 &.&.&.&.&.&.\\
1&0&.&.&.&.&.\\.&1&0&.&.&.&.\\.&.&1&0&.&.&.\\.&.&.&.&.&.&.\\.&.&.&.&.&.&.\\
.&.&.&.&.&.&.
\\ \end{array} \right) \hskip 1cm
D_i = \left( \begin{array}{lllllll} d_i &{d_i\over v_i}&{d_i\over v_i^2}&
{d_i\over v_i^3}&{d_i\over v_i^4}&.&.\\
0&1&{1\over v_i}&{1\over v_i^2}&{1\over v_i^3}&.&.\\0&0&1&{1\over v_i}&
{1\over v_i^2}&.&.\\.&.&.&.&.&.&.\\.&.&.&.&.&.&.\\.&.&.&.&.&.&.\\.&.&.&.&.&.&.
\\ \end{array} \right) \ee
{\bf Remarks}:\\ \\ 1) If $ \epsilon = 0 $,then all the $ d_i = 1 $ and using (50)
one sees
that $ D_i D_j = D_j D_i \ \ \ \forall i , j $ although $ D_i E \ne E D_i $.
Hereafter we restrict ourselves this case. Note that this does not mean that
the two point functions $ < \tau^{(i)}_k  \tau^{(j)}_l > $\  and\  $< \tau^{(j)}_k  \tau^{(i)}_l  > $
are equal.\\ \\
2) When $ \epsilon = 0 $ we have $ D_i = D_j $ for $ v_i = v_j $, so that
equation (7) becomes vaccuous in this case and no singularity arises due
to $ v_i = v_j $ in this equation. Furthermore one can safely eliminate
one of these generators for the other one. This means that the (p-1)-ASEP
algebra is naturally embedded in the (p)-ASEP algebra.\\
The ket and bra vectors $ \vert V> $ and $ < W \vert $ are found to be:
\be \vert V > = \sum_{n=0}^{\infty}( 1-\beta)^{n}\vert n > \ee
\be < W \vert = \sum_{n=0}^{\infty}( \alpha)^{-n}<n \vert \ee

Having a representation at hand is only half of the way in obtaining explicit
expressions for the physical quantites like currents and correlation functions.
To obtain these quantities,one should do lenghty calculations on the matrix
elements
of porduct of operators.Our work is not complete in the sence that we do not
yet have the final explicit expression for these matrix elements. However we
explore the properties of the representations as far as we can, hoping
that using these,
the rest of the problem be solved in another occasion, by this author or
by others.
\subsection{Eigenvectors of\ \ $ D_i$\ 's}

Let $ z $ be any complex number, with $ \vert z \vert < v_1 $. Define:
\be \vert z > = \sum_{n=0}^{\infty} z^{n}\vert n > \ee
Then $ \vert z > $ is a common eigenvector of all $ D_i $'s:
\be D_i \vert z > = { v_i\over { v_i - z}}\vert z > \ee
This can be easily proved by direct calculation. The states $ \vert z > $ can in
fact be thought of as the coherent states of the algebra constructed as follows:
$ \vert z > := { 1 \over { 1 - z E }}\vert 0 >$
where $ {1\over { 1-z E}}$ is understood by its power series.
We now define the dual vector $ < z \vert = \sum_{n=0}^{\infty} z^{-n}<n\vert $.
Then we have the following theorem, the proof of which is accomplished
by noting that
$ <n\vert E = < n-1 \vert $ \ \ and \ \ $<0\vert E = 0 $, and doing
straightforward calculations.\\ \\
{\bf Theorem}:\\ \\
{\bf a})$ <z\vert E = {1\over z } <z \vert $\\
{\bf b}) The coherent states form a basis of the space W, and:
$$\vert n > = \int_{\vert z\vert \leq 1} {1\over \pi} dz d{\bar z}
z^{-n}\vert z > $$
{\bf c}) The coherent states $ \vert z > $ form an over complete basis for
W. A complete
basis is obtained by taking only the states with fixed $ \vert z \vert $,
with a completeness relation:
$$ 1 = \oint_{fixed \vert z \vert} {d z \over {2 \pi i z} }\vert z ><z\vert $$
{\bf d}) $ <\omega\vert z > = {1\over {1-{z\over \omega}}} \ \ $ for \ \
$ \vert z\vert < \vert \omega \vert$\\ \\
{\bf Remarks}:\\
1) It is now appropriate to recall that the states $ \vert V> $ and $< W\vert $ are
in fact coherent states and so it is best to denote them respectively by
$ \vert 1-\beta> $ and$ < \alpha \vert.$\\
2) The calculation of any physical quantity is now reduced to the
calculation of
matrix elements of powers of $ C $ between the coherent states. For example,
the average density $ < n^{i}_k>$ is given as:
\be  < n^{i}_k> ={1\over Z_N}  \oint_{(1-\beta < \vert z\vert < \alpha,\ \
\vert z\vert < v_i)}{d z \over {2 \pi i z} }
<\alpha\vert C^{k-1}\vert z><z\vert C^{N-k}\vert 1-\beta>
{ v_i\over { v_i - z}}\ee
where $ Z_N = <\alpha \vert C^N \vert 1-\beta > $.
Or the probability of a segment [k,l] be empty is given by
\be  P(k,l):={1\over Z_N} \oint_{1-\beta < \vert z\vert < \alpha,}{{d z}
\over {2 \pi i z} }
<\alpha\vert C^{k-1}\vert z><z\vert C^{N-k-l}\vert 1-\beta>
{ 1\over z^{l-k}}\ee

Once the eigenvalues and eigenvectors of $ C $ are known,the problem can
be solved completely.
In the following we will give an implicit formula for these objects.\\\\
\subsection{Eigenvalues of C}

First we consider the case $ p=1$.
For any coherent state $ \vert z > $ we have $ E \vert z > = {1\over z}
( \vert z> - \vert 0 > ) $.
For any two complex numbers $ z $ and $ \omega $ define:
\be \vert z,\omega > = z \vert z> - \omega \vert \omega > \ee
Then it follows that:
\be E\vert z,\omega > = \vert z> - \vert \omega > \ee
Since for $ p=1$ we have $ D \vert z > = {1\over { 1-z}}\vert z > $ we obtain:
\be C\vert z,\omega > = (D+E) \vert z, \omega>={1\over { 1-z}} \vert z>-{1\over {1-\omega}} \vert \omega > \ee
Therefore the state $ \vert z, \omega> $ is an eigenstate of $ C $ if we have
$ z(1-z)=\omega(1-\omega)$, the solution of which is $\omega =z$ \ or $\omega = 1-z$
The second solution is acceptable and hence we have:
\be C\vert z,1-z > = {1\over {z(1-z)}} \vert z,1-z> \ee
Now let $ p>1$. Using (56) and (60), we obtain:
\be C \vert z,\omega > = z \eta (z) \vert z> - \omega \eta (\omega)
\vert \omega > \ee
where
\be \eta(z) = {1\over z} + {1\over p}\sum_{i=1}^p {v_i\over {v_i-z}}\ee
The state $ \vert z,\omega > $ is an eigenvector of $C$ if $ z $ and $ w $ lie
on the curve $ \eta(z) = \eta(\omega) $. We will then have:
$C\vert z, \omega(z)> = \eta( z )\vert z, \omega(z)>$. \\
Note that in the $ p\longrightarrow \infty$ $ z $ and $ \omega $ are
related by the following equation:
\be {1\over z} + \int { v P(v) \over {v-z}} =  {1\over \omega} + \int { v P(v) \over {v-\omega}} \ee

In order to understand the connection with the eigenstate obtained in
Derrida et al [2] in the $p=1$ case we change the basis of W as:
\be \vert e_n > = ( E-1)^n \vert 0 > ,\ \ \ \ \vert e_n > = \vert 0> \ee
Using ( 6,7) it is easily found that:
$$ E\vert e_n > = \vert e_n > + \vert e_{n+1}> $$
$$ D\vert e_n > = \vert e_n > + \vert e_{n-1}> $$
$$ C\vert e_n > = \vert e_{n-1}>+ 2\vert e_n> +\vert e_{n+1}>$$
Thus we have
\be \vert z > = { 1 \over { 1 - z E }}\vert 0 >= { 1 \over { 1-z-z(E-1)}}
\vert 0 >= { 1 \over { 1 - z  }}\Bigg({ 1 \over { 1 -{z\over { 1-z}} E }}
\Bigg)\vert 0 >={ 1 \over { 1 - z }}\vert {z\over {1-z}} >_d\ee
where by $ \vert z>_d $ we mean $ \vert z >_d =  \sum_{n=0}^{\infty} z^{n}\vert e_n > $
Therefore
\be \vert z,1-z>=
{ z \over { 1 - z }}\vert {z\over {1-z}} >_d
-{{1-z} \over z}\vert {{1-z}\over z} >_d  =
 \sum_{n=0}^{\infty} \bigg ( ({z \over { 1 - z }})^{n+1} -
({{1-z}\over z})^{n+1}
\bigg)\vert e_n > \ee
Taking $ {z\over {1-z}}= e^{i\theta}$ gives
\be \vert z , 1-z > = \sum_{n=0}^{\infty} sin ( n+1)\theta \vert e_n > \ee
which is denoted by $ \vert \theta > $ in [2] with eigenvalue $
{1\over z(1-z)} = 2 ( 1+cos \theta) $.
This concludes our treatment of the representations of the algebra.

\section{Discussion}

This work can be pursued further in the following directions:\\
a) Finding a solution of the equation $ \eta(z) = \eta(\omega) $ either for
a low value of $ p \ ( e.g. p = 2 )$ or in the $ p\longrightarrow
\infty$ limit, and then expanding the state $ \vert 1-\beta > $
in terms of the eigenstates of $ C $. In this way one will obtain the current
and the phase diagram of the system.\\
b) Finding a solution of the mean field equations, either numerically or
analytically. These equations can be written down from (31). In the $
p\longrightarrow \infty$
limit they are:
$$ \alpha v e_1 = J_k(v) = J_{k+1}(v) = ... = ( v+\beta - 1 ) n_N (v) $$
where
$$ J_k (v) = v n_k(v)e_{k+1} + \int_{0}^{v}(v-v')n_k(v)n_{k+1}(v')dv'
-\int_{v}^{\infty}(v'- v)n_k(v')n_{k+1}(v)dv'$$
Here $ n_k(v) $ is average density of particles of speeds $ v $ at site $ k $
and $ e_k := 1-\int_{0}^{\infty}n_k(v) dv $ is the probability of site
$k$ being vacant. In the one species model it is known that the mean field
analysis gives the phase diagram correctly. So it will be interesting to see
for a typical probability distribution how the phase diagram will
be modified.\\  \\
{\bf Acknowledgements}\\ \\
My Special Thanks go to A.Langari for helping me in drawing the figures.
I would also like to thank  R. Ejtehadi, J. Davoodi,and M.R. Rahimitabar for stimulating discussions,
and F. Jafarpour,
M. E. Fouladvand and A. Shafi Dehabad for valuable comments.
\newpage
{\bf Figure Captions}\\    \\
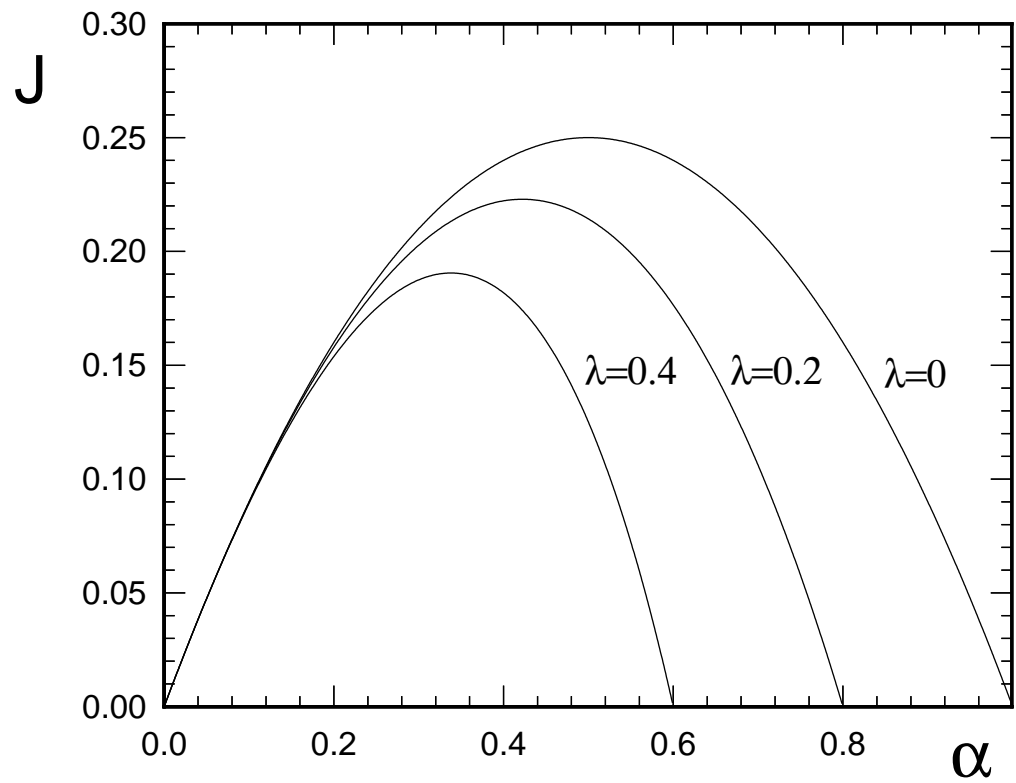
Figure 1: The current versus the arrival rate of particles for different
values of ${\lambda}$.\\
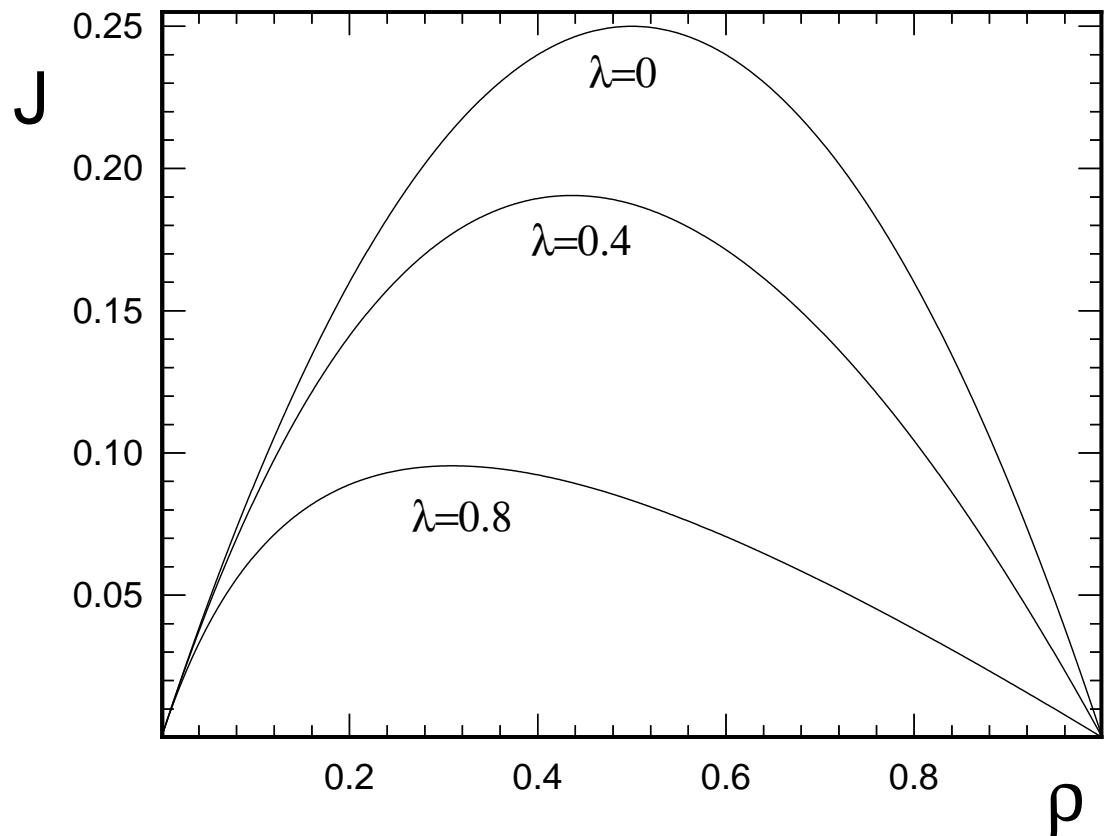
Figure 2: The current versus the density for different widths of
the distribution.\\
\newpage
\large {\bf References}
\begin{enumerate}

\item Hakim V and Nadal J P 1983 J.Phys. A ; Math. Gen. {\bf 16} L213
\item Derrida B Evans M R Hakim V and Pasquier V 1993 J.Phys.
A ; Math. Gen. {\bf 26} 1493
\item Liggett T M 1985 Interacting Particle Systems ( New York; Springer)
\item Derrida B 1998 Phys. Rep. {bf 301} 65
\item Janowsky S A and Lebowitz J L 1994 J. Stat. Phys. {\bf 77} 35
\ \ \ Janowsky S A and Lebowitz J L 1992 Phys. Rev. A {\bf 45} 618
\item Derrida B Janowsky S A Lebowitz J L and Speear E R 1993 Europhys
. Lett.{\bf 22} 651
\\ \ \ Derrida B Janowsky S A Lebowitz J L and Speear E R 1993
J. Stat. Phys.{\bf 73} 831
\item Mallik K 1996 J. Phys. A ; Math. Gen. {\bf 29} 5375
\item Evans M R  Foster D P Godreche C and Mukamel D 1995 J. Stat. Phys. {\bf
80} ; Phys. Rev. Lett.{\bf 74},208(1995).
\item Hinrichsen H 1996 J.Phys. A ; Math. Gen. {\bf 29} 3659
\item Rajewsky N Schadschneider A and Schreckenberg M 1996 J.Phys.
A ; Math. Gen. {\bf 29} L305
\item Honecker A and Peschel I 1996 J.Stat. Phys. {\bf 88} 319
\item Schreckenberg M Schadschneider A Nagel K and Ito M 1995
Phys. Rev. E {\bf 51} 2339
\item Lee H W Popkov V and Kim D 1997 J.Phys. A ; Math. Gen. {\bf 30} 8497
\item Evans M R 1996 Europhys. Lett. {\bf 36}13
\item Kreb K and Sandow S 1997 J.Phys. A ; Math. Gen. {\bf 30} 3165
Stochastic Systems and Quantum Spin Chains {\it preprint} Cond-mat/9610029
\item Arndt P Heinzel T and Rittenberg V, 1998 J.Phys. A ; Math. Gen. {\bf 31} 833
\item Alcaraz F C Dasmahapatra S and Rittenberg V, 1998 J.Phys. A ; Math. Gen. {\bf 31} 845

\end{enumerate}

\end{document}